\documentclass{article}\usepackage[]{graphicx}\usepackage[]{color}
\makeatletter
\def\maxwidth{ %
  \ifdim\Gin@nat@width>\linewidth
    \linewidth
  \else
    \Gin@nat@width
  \fi
}
\makeatother

\definecolor{fgcolor}{rgb}{0.345, 0.345, 0.345}

\usepackage{framed}
\makeatletter
 {\par\unskip\endMakeFramed%
 \at@end@of@kframe}
\makeatother

\definecolor{shadecolor}{rgb}{.97, .97, .97}
\definecolor{messagecolor}{rgb}{0, 0, 0}
\definecolor{warningcolor}{rgb}{1, 0, 1}
\definecolor{errorcolor}{rgb}{1, 0, 0}

\usepackage{alltt}
\usepackage{authblk}

\title{Opinion Dynamics and Collective Decisions}

\author{Jan Lorenz}
\affil{Department of Psychology and Methods, Jacobs University, Bremen, Germany; and \\
 ComputationalSocial Science Department, GESIS Leibniz-Institut f\"ur Sozialwissenschaften, Cologne, Germany. \\
 post@janlo.de}
\author{Martin Neumann}
\affil{Johannes-Gutenberg Universit\"at, Mainz, Germany. m.neumann@uni-mainz.de}

\date{Preprint for Editorial for a Topical Issue in \emph{Advances in Complex Systems} 2018}
\IfFileExists{upquote.sty}{\usepackage{upquote}}{}
\begin{document}
\maketitle
\begin{abstract}
We expect that democracy enables us to utilize collective intelligence 
such that our collective decisions build and enhance social welfare, and 
such that we accept their distributive and normative consequences. 
Collective decisions are produced by voting procedures which aggregate individual preferences and judgments. 
Before and after, individual preferences and judgments change 
as their underlying attitudes, values, and opinions change through discussion and deliberation. 
In large groups, these dynamics naturally go beyond the scope of the individual
and consequently might show unexpected self-driven macroscopic systems dynamics following socio-physical laws.
On the other hand, aggregated information and preferences as communicated through 
media, polls, political parties, or interest groups, also play a large role in the individual opinion formation process. 
Further on, actors are also capable of strategic opinion formation 
in the light of a pending referendum, election or other collective decision.
Opinion dynamics and collective decision should thus not only be  
tackled by social choice, game theory, political and social psychology, 
but also from a systems dynamics and sociophysics perspective.
\end{abstract}


In the year 2000, a topical issue on Simulation in the Social Sciences appeared in this journal \cite{2000IntroductionWhySimulation}. In this topical issue, one of the most influential articles on opinion dynamics by \emph{Deffuant et al} \cite{Deffuant.Neau.ea2000MixingBeliefsamong} appeared. Since then, opinion dynamics has emerged as one of the major topics in computational social science. This topical issue takes up this research field and broadens the perspective by putting opinion dynamics into the wider framework of political science. Namely, the problem of collective decision making. 

The problem of collective decision making exists in two versions, as a question of emergence and as the problem of aggregation. The first often runs under the keyword ``opinion dynamics'' and is a typical problem in the science of complex systems, showing how different landscapes of political attitudes such as a polarized or consensual one may emerge. The second version is how preferences of individuals should be aggregated to a collective one. This social choice problem is one of the core problems in political science. Complex systems researcher often focus solely on the emergence of collective behavior from repeated decentralized interaction of many individuals and neglect the existence of aggregating institutions. On the other hand, research on the design of aggregating institutions often put less attention to societal processes happening before aggregation. This topical issue puts together new insights on different facets of this nexus of opinion dynamics and collective decision.

\section{Collective decisions}

Voting procedures which aggregate a collective preference relations from several individual preference relations intrinsically create obstacles to achieve their democratic aim of properly aggregating people's will. 
Social choice theory characterizes the possibilities to aggregate a collective decision from individual preferences. Under the condition of non-dictatorship a well-defined aggregation procedure exists only when either it is simply dichotomous (Yes/No or one policy dimension); or if preference heterogeneity is low such that properties of the dichotomous case are resembled; or when some individuals have veto power over all collective decisions \cite{Austen-Smith.Banks1999PositivePoliticalTheory}. This resembles several results in Social Choice Theory, namely Arrow's impossibility theorem \cite{Arrow1951SocialChoiceand}, the Gibbard-Satterthwaite theorem \cite{Satterthwaite1975Strategy-proofnessandArrows,Gibbard1973ManipulationofVoting} about the ubiquity of tactical voting, the discursive dilemma \cite{Pettit2001DeliberativeDemocracyand} and Condorcet's paradox \cite{Condorcet1785Essaisurlmapplication}. Further on, majority rule is the only reasonable rule which treats voters and respectively alternatives equally \cite{May1952SetofIndependent}. In a more-dimensional policy space the ``chaos'' theorem \cite{McKelvey1976Intransitivitiesinmultidimensional} shows that one can steer the collective decision to any point by sequential majority votes. These results make the design of aggregation procedures an intrinsically delicate endeavor. 

The paradoxes, impossibility results and dilemmas of social choice theory are rephrased by the \emph{democratic trilemma} \cite{List2011LogicalSpaceof}, which says that any aggregation rule can only fulfill two of three principles: 
\begin{itemize}
\item robustness to pluralism (all preference relations are allowed as individual input), 
\item basic majoritarianism (nothing shall be implemented when a majority prefers something else), and 
\item collective rationality (aggregated preferences are transitive and complete). 
\end{itemize}

Consequently, routes out of the democratic trilemma are possible by relaxing one of these desired principles. We skip the idea to relax collective rationality here and focus first on relaxing basic majoritarianism. That means that a collective decision can be made by a procedure, even when there is another possible decision which would win against it by a majority vote. One procedure is a terminating chain of dichotomous majority votes, where the decision about the proposals or about the order of votes is externalized to another entity (the agenda-setter) or a different procedure. \cite{Krause1995Essentiallylexicographicaggregation} showed that an essentially lexicographic order of importance in policy dimensions is natural in multi-criteria decision making. This does not seem appropriate for all kinds of decisions, e.g. not for budgeting when there is no seniority of budget items. An endogenous way of relaxing majoritarianism is to find a suitable and acceptable procedure of aggregation which defines a ``best fitting'' collective outcome. 
Considering the distribution of money among three projects, a possible procedure of aggregation would be that each agent hands in a proposal and the average of all three proposals is implemented. Weighted arithmetic averaging is the only procedure for this purpose which fulfills the conditions of zero unanimity and neutrality of projects \cite{Lehrer.Wagner1981RationalConsensusin}. 

Interestingly, averaging and similar aggregation procedures are much less studied and rarely used in reality, although they provide a way out of the democratic trilemma. Consequently, it is of particular interest to understand how to design them in a fair way, in particular with respect to how voting weights correspond to power. In this topical issue, \emph{Sascha Kurz} \cite{Kurz2018Importanceinsystems} makes some first steps to mathematically analyze this measurement of importance in systems with interval decisions. He generalizes the definitions of two famous power indices, the Shapley-Shubik and Penrose-Banzhaf index, from binary voting to procedures which aggregate real values or vectors of real values. 

The last route out of the democratic trilemma is to relax robustness to pluralism. In a multidimensional space of possible decisions, that means to restrict the eligible input. If this is not enforced by the procedure itself this puts the burden back on voters to deliver individual preferences for which aggregation can stand majority votes. This is most easily fulfilled when individuals have a consensus about collective preferences. DeGroot \cite{DeGroot1974ReachingConsensus} and Lehrer and Wagner \cite{Lehrer.Wagner1981RationalConsensusin} pioneered the study under which conditions consensus evolves through repeated averaging. 

In this topical issue, \emph{Ulrich Krause} \cite{Krause2018DynamicalModelProcess} builds on this model and investigates the question if and how a process of repeated sharing of wealth within a group leads to a more equal distribution. For this purpose, a formal model of a matrix of sharing rates within a group is developed and its mathematical properties are analyzed. He proves that, in fact, the limiting distribution of wealth is equality. Given the condition of a mathematical ring structure this finding holds for surprisingly many conditions such as uniform, general or even varying sharing rates. 

If people are not able to produce consensus, deliberation can still promote the achievement of a metaconsensus \cite{Dryzek.List2003SocialChoiceTheory}, where individuals agree in a way such that remaining alternatives form an ordered set (as the political left---right continuum) from which the electorate can then stably decide on for the median voter's choice \cite{Hotelling1929StabilityinCompetition,Black1948RationaleofGroup,Downs1957BookEconomicTheoryof}. This can be generalized to alternatives which form a partially ordered set (a lattice) \cite{Pivato2009Geometricmodelsof}. 
The hope is thus, that deliberation causes preference changes of individuals such that the ``relevant'' dimensionality of the space of collective decisions is reduced to be ``essentially'' one. 
%

\section{Preference and belief formation} 

How individuals form preferences and beliefs before they are put into a procedure of aggregation is usually left out by social choice theory, although the political process we experience in reality is a never ending stream of opinions uttered publicly and privately by a multitude of individuals for a multitude of purposes. For the theory of deliberative democracy \cite{Bohman.Rehg1997DeliberativeDemocracyEssays,Mutz2008IsDeliberativeDemocracy} this stream should ideally be a process of transformation of individual preferences through rational discussion towards a consensus about the common good \cite{Mansbridge1983BeyondAdversaryDemocracy,Habermas.Rehg1998BetweenFactsand,Elster1997MarketandForum}. Although people are able to deliberate \cite{Steenbergen.Bachtiger.ea2003MeasuringPoliticalDeliberation,Baechtiger.Niemeyer.ea2010DisentanglingDiversityin}, there is evidence that this can moderate \cite{Meffert.Guge.ea2004Goodbadand} but also polarize views  
\cite{Wojcieszak2011DeliberationandAttitude}. 
Moreover, depolarization happens usually only on factual and much less on value-laden issues \cite{Vinokur.Burnstein1978Depolarizationofattitudes}.
In deliberative opinion polling \cite{Fishkin.Luskin2005ExperimentingwithDemocratic} people indeed change opinions and this can lead to a reduction of majority cycles \cite{List.Luskin.ea2006DeliberationSinglePeakedness} but rarely on salient topics \cite{Farrar.Fishkin.ea2010DisaggregatingDeliberationsEffects}. 
\cite{Mansbridge1983BeyondAdversaryDemocracy} argues that groups with mostly common interests can benefit from deliberation, while groups with mostly conflicting interests might suffer. 

A core questions is thus, under what conditions an interacting crowd ultimately using an aggregating institution is able to  elicit the wisdom of the crowd \cite{Surowiecki2004wisdomofcrowds} rather than the madness of the crowd \cite{Mackay1848MemoirsofExtraordinary}. \emph{Christan Ganser} and \emph{Marc Keuschnigg} tackle this question in this topical issue by analyzing how  social influence changes the wisdom of the crowd \cite{Ganser.Keuschnigg2018SocialInfluenceStrengthens}. They use a lens model where individuals aggregate several cues to a score value for each alternative. These scores could then be subject to opinion dynamics before they are aggregated to a group decision for one alternative, either through a voting approach or an averaging approach. It turns out that under the voting procedure, social influence improves the accuracy of collective decisions, while it does not under the averaging procedure.

\section{Opinion dynamics in larger groups}

In larger groups, even rational players can not focus on the opinions and preferences of all other individuals and compute game theoretic equilibria. They have to base their decision on individual interaction, knowing that there are unknown others who form and change the opinion landscape at the same time. 
Therefore, it is important to understand opinion dynamics also on a larger scale, going beyond the scope of the individual. 
Landscapes of values, preferences and political positions appear fragmented often as long-lasting cleavages \cite{Lipset.Rokkan1967PartySystemsand} or systemic polarization \cite{Pardos-Prado.Dinas2010Systemicpolarisationand}. Such opinion landscapes can be explored by agent-based simulation models where citizens adapt and adjust their opinions through communication. 
Already in the year 2000, Krause \cite{Krause2000DiscreteNonlinearand} extended the simple linear model with fixed  weights for averaging \cite{DeGroot1974ReachingConsensus} by the bounded confidence assumption, saying that individuals assign positive weights only to individuals which are not too far away in the opinion space. Together with Rainer Hegselmann, they showed by simulation \cite{Hegselmann.Krause2002OpinionDynamicsand} that fragmented opinion distributions with local consensus in subgroups can emerge in that way. It turned out, that also the core assumption in the opinion dynamics model of Deffuant et al from the year 2000 \cite{Deffuant.Neau.ea2000MixingBeliefsamong} is bounded confidence. Coming from quite different backgrounds, these two papers were written independently of each other, but triggered a fruitful stream of simulation studies in opinion dynamics \cite{Flache.Maes.ea2017ModelsSocialInfluence}, about systems dynamics towards consensus, polarization, fragmentation, or extremism.

In this topical issue \emph{Sylvie Huet} and \emph{Jean-Denis Mathias} \cite{Huet.Mathias2018Fewselfinvolved} use the framework of opinion dynamics to study the emergence of norms, which they see as majority's behavior or opinion. Thus, they start from opinion dynamics models which they extended with several components drawn from social-psychology. They differentiate between ordinary and self-involved agents, representing actors who are particularly engaged in certain issues. Self-involved agents are characterized by a main and a complementary attitude. If two self-involved agents are close on the main attitude they get attracted also on the complementary attitude. If they disagree on the main component they shift away from each other also in the complementary one. In contrast the opinion dynamics of ordinary agents follows the dynamics known from the opinion dynamics tradition. Thus, Huet and Matthias study a two dimensional opinion space with varying constellations. This setting generates various dynamics: Agents may simply converge to a central consensus. However, consensus on one dimension might also lead to polarization on the other dimension. This formal framework enables a discussion of potential mechanisms of norm emergence. 

\emph{Guillaume Deffuant}, \emph{Ilaria Bertazzi}, and \emph{Sylvie Huet} \cite{Deffuant.Bertazzi.ea2018darksidegossips} introduces an opinion dynamics model extended such that agents also have opinions on themselves and on others. The model follows the assumption of opinion dynamics models that during interactions the agents change opinions towards the opinion of the other agent. Gossip is introduced as agents changing not only their opinion about themselves but also about random other agents.  The model is run with and without gossip. Starting from opinions at zero, without gossip the average opinion starts to drift towards a positive value. However, introducing gossip changes the dynamics: starting from zero the average opinion drifts towards a negative value. This is called the dark side of gossip and is caused by the relative influence of a positive bias on self-opinions and of a negative bias on opinions about others. Gossips increases the negative bias about others, which can dominate the positive bias on self-opinions, leading to a negative average opinion.

Also \emph{Andreas Flache} \cite{Flache2018Aboutrenegadesand} studies opinion dynamics where attitudes about others play a role but on the aggregated level of intergroup attitudes which are themselves also subject to social influence. 
In particular, a renegade minority of outgroup lovers can play a key role in avoiding mutually negative intergroup relations
and even elicit reversed polarization, resulting in a majority of individuals developing a negative attitude towards their ingroup and a positive one of the outgroup.

\section{Spreading in networks and segregation}

Other aspects of the nexus of opinion dynamics and collective decisions is how rumors or other information spreads or does not spread through a social network, and how different homophily preferences of individuals trigger the emergence of attitude segregation. 

\emph{Amirhosein Bodaghi} and \emph{Sama Goliaei} study rumor spreading within online social networks inspired by twitter data \cite{Bodaghi.Goliaei2018NovelModelRumor}. They go beyond early models that represent rumor spreading by a mechanism of epidemic contagion by including more realistic socio-psychological theories. Specifically, the model includes an effect of dissenting opinions on rumor spreading behavior. The hypothesis is formulated as a stochastic model that is analyzed with simulation. For testing their hypothesis, they compare it with a baseline model without the hypothesis. The results show that including the hypothesis is better capable to replicate the empirical data.

\emph{Mehrdad Kermani} and \emph{Reza Ghesmati} and \emph{Masoud Jalayer} study influence maximization \cite{Kermani.Ghesmati.ea2018OpinionAwareInfluence} which is problem to find a small subset of seed nodes to maximize the diffusion or
spread of information. In their paper, they extend the baseline model of information spreading by giving individuals opinion and the message an opinionated content. Thus, they stud opinion aware influence maximization. In this problem, the
main objective is to maximize the spread of a desired opinion, by optimizing the
message content, rather than the number of infected nodes. Experimental results on
some of the well-known datasets show the efficiency and applicability of their proposed
algorithm.

\emph{Rocco Paolillo} and \emph{Jan Lorenz} extend Schelling's model of ethnic segregation \cite{Schelling1971Dynamicmodelsof} to study how different homophily preferences mitigate and spur ethnic and value segregation
\cite{Paolillo.Lorenz2018Howdifferenthomophily}. They take into account that in modern, ethnically diverse societies people might not define similarity based on ethnicity but many rely on shared tolerance towards ethnic diversity. The introduction of such value-oriented agents reduces total ethnic segregation when compared to Schelling's original model, but the new phenomenon of value segregation appears to be more pronounced than ethnic segregation. Furthermore, due to cross-contagion, stronger ethnic homophily preferences lead not only to greater ethnic segregation but also to more value segregation. Stronger value-orientation of the tolerant agents similarly leads to increased ethnic segregation of the ethnicity-oriented agents.

\section{Conclusion}

The collection of the papers in this topical issue brings together the different topics of preference formation and aggregation of preferences into a collective decision. The papers by Huet et al., Deffuant et al. and Flache highlight how the research field of opinion dynamics has diversified by taking more fine grained assumptions into account. Kurz and Krause study properties of aggregation procedures and collective outcomes of a bottom-up process of sharing. Ganser and Keuschnigg analyse conditions of collective intelligence whereas Kermani et al. and Paolillo and Lorenz study properties of networks. The latter paper builds on and extends Schelling’s well known model of segregation. All contributions have in common that they utilize a mathematical and computational approach for studying complex systems of multiple interacting agents. Thereby this topical issue highlights that the analysis of systems of multiple components benefits from cross-disciplinary approaches and findings of system dynamics and socio-physics. This approach picks-up the theme of the topical issue on simulation in the social sciences in the year 2000 and demonstrates the progress that has been made since these times. The ongoing research demonstrate the potential of a formal analysis of political culture and democratic decision making and we hope that the topical issue will stimulate further debates and future research as it has been the case for the topical issue nearly 20 years ago, last but not least resulting in this topical issue.

\section*{Acknowledgments}
The idea for this topical issue was kicked off at the ``Interdisciplinary Workshop on Opinion Dynamics and Collective Decision 2017'' ODCD2017, July 5-7, 2017 at Jacobs University Bremen, Germany. It received substantial funding from the 
German Research Foundation (Deutsche Forschungsgemeinschaft DFG, Research Project ``Opinion Dynamics and Collective Decision'' Grant number 265108307). 

\vspace*{-3pt}   


\bibliographystyle{ws-acs}


\end{document}